\documentclass[journal=jctcce,manuscript=article,
layout=twocolumn]{achemso}

\usepackage{chemformula} 
\usepackage[T1]{fontenc} 

\usepackage{graphicx}
\usepackage{dcolumn}
\usepackage{amsmath}
\usepackage{amssymb}
\usepackage{mathtools}
\usepackage{txfonts}
\usepackage{url}
\usepackage{multirow}
\usepackage{here}
\usepackage{lscape}
\usepackage{braket}

\usepackage{booktabs}
\usepackage{multirow}

\usepackage{hyperref}
\usepackage{physics}
\usepackage{longtable}

\makeatletter
\def\Hline{%
\noalign{\ifnum0=`}\fi\hrule \@height 1pt \futurelet
\reserved@a\@xhline}
\makeatother

\author{Abhishek Raghav}
\email{mwkabr1915@icloud.com}
\affiliation[SISSA]
{International School for Advanced Studies (SISSA), Via Bonomea 265, 34136, Trieste, Italy}

\author{Ryo Maezono}
\affiliation[JAIST]
{School of Information Science, Japan Advanced Institute of Science and Technology (JAIST), Asahidai 1-1, Nomi, Ishikawa 923-1292, Japan}

\author{Kenta Hongo}
\affiliation[JAIT]
{Research Center for Advanced Computing Infrastructure, Japan Advanced Institute of Science and Technology (JAIST), Asahidai 1-1, Nomi, Ishikawa 923-1292, Japan}

\author{{\color{black}Sandro Sorella}}
\affiliation[SISSA]
{International School for Advanced Studies (SISSA), Via Bonomea 265, 34136, Trieste, Italy}

\author{Kousuke Nakano}
\email{kousuke_1123@icloud.com}
\affiliation[NIMS]
{Research and Services Division of Materials Data and Integrated System, National Institute for Materials Science (NIMS), Tsukuba, Ibaraki 305-0047, Japan}

\title {Towards chemical accuracy using the Jastrow correlated antisymmetrized geminal power \textit{ansatz}}

\begin{document}

\begin{tocentry}
\centering
 \includegraphics[width=\linewidth]{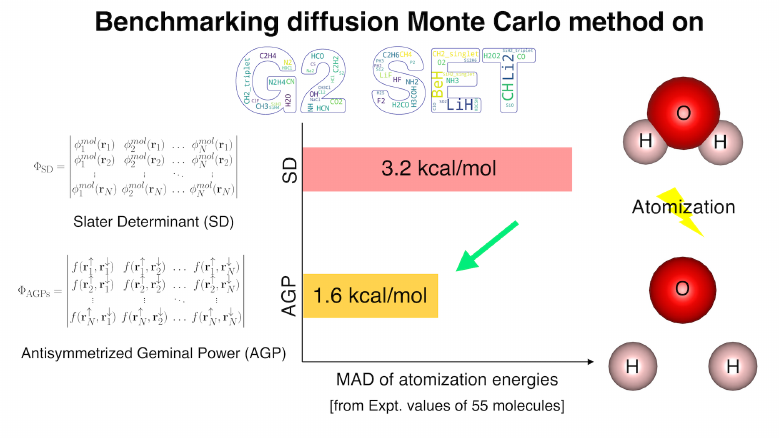}
 \label{For Table of Contents Only}

\end{tocentry}

\begin{abstract}
Herein, we report accurate atomization energy calculations for 55 molecules in the Gaussian-2 (G2) set using lattice regularized diffusion Monte Carlo (LRDMC). 
We compare the Jastrow--Slater determinant \textit{ansatz} with a more flexible JsAGPs (Jastrow correlated antisymmetrized geminal power with singlet correlation) \textit{ansatz}. 
AGPs is built from pairing functions, which explicitly include pairwise correlations among electrons and hence, this \textit{ansatz} is expected to be more efficient in recovering the correlation energy. 
The AGPs wave functions are first optimized at the variational Monte Carlo (VMC) level, which includes both the Jastrow factor and the nodal surface optimization. 
This is followed by the LRDMC projection of the \textit{ansatz}. 
Remarkably, for many molecules, the LRDMC atomization energies obtained using the JsAGPs \textit{ansatz} reach chemical accuracy ($\sim$1 kcal/mol) and for most other molecules, the atomization energies are accurate within $\sim$5 kcal/mol. 
    We obtained a mean absolute deviation of 1.6 kcal/mol with JsAGPs and 3.2 kcal/mol with JDFT (\textbf{J}astrow factor + Slater determinant with \textbf{DFT} orbitals) \textit{ansatz}.
This work shows the effectiveness of the flexible AGPs \textit{ansatz} for atomization energy calculations and electronic structure simulations in general.
\end{abstract}



\section{Introduction}
\textit{Ab initio} quantum Monte Carlo (QMC) techniques~\cite{2001FOU}, such as variational Monte Carlo (VMC) and diffusion Monte Carlo (DMC) have been used to compute accurate many-body wavefunctions (WF) for atomic, molecular, simple crystal, and even complex electronic systems~{\cite{2020NEE, 2020KEN, 2020NAK}}. 
Benchmark results are of great interest for QMC methods because these methods have the capability to match the accuracy of quantum chemical methods and also scale more favorably than several wave function based methods~\cite{2001FOU}.
A WF-based quantum chemical method, such as the coupled-cluster with single and double and perturbative triple excitations (CCSD(T)), scales as $O(N^7)$, where \textit{N} is the number of 
electrons~\cite{2006FEL,2021GYE}.
%
In contrast, DMC at most scales as $O(N^4)$ and is closer to the computational scaling of $O(N^3)$ (for $N < 1000 \sim 2000$)~\cite{2020NEED} of the more traditional mean-field methods like DFT. Since there is a large prefactor involved in the DMC method, the WF-based method is often more favorable for smaller systems than DMC. 
However, as the system size increases, the scaling becomes the dominating factor in their computational costs. 
It limits the application of the \textit{conventional} CCSD(T) to systems of up to $\sim$80-90 electrons~\cite{2013RAM}, while recent implementations extend this limit further, e.g., up to $\sim$300 electrons for molecules~\cite{2021NAG} and $\sim$100 electrons for solids.~\cite{2018TSA} 
On the other hand, for QMC methods, system sizes up to $\sim$2000 electrons are accessible, which 
makes them invaluable for condensed matter calculations~\cite{2020NEED}.
Thus, making QMC methods more accurate and efficient using various approaches (like using more
flexible WF \textit{ansatz}) is an active area of research.

\vspace{2mm}
A widely used set of benchmark data is the atomization energies of the Gaussian-2 (G2) set of molecules~\cite{1991CUR}.
It contains 55 small molecules composed of elements from the first, second, and third rows of the periodic table. 
Accurate experimental values for the atomization energies of the G2 set of molecules are available, which makes it very attractive for benchmarking.
Another way to benchmark is to compare the computed total energies with the estimated exact energies. 
However, to evaluate the error cancellation, it is important to benchmark the atomization energies.

\vspace{2mm}
The G2 set benchmarks have been used to test several state-of-the-art \textit{ab initio} computational methods.
Chemical accuracy (deviation of 1 kcal/mol from experimental estimates of atomization energies) has often been used as the target accuracy. 
The benchmarking relative to molecular atomization energies of DFT methods in previous studies resulted in a relatively large mean absolute deviation (MAD) value of $\sim$40 kcal/mol for local density approximation (LDA) and $\sim$2.5 kcal/mol for the hybrid B3LYP functional.~\cite{1995BAU}.
In another study, B3LYP on an extended G2 set gave an MAD of 3.11 kcal/mol~\cite{1997CUR}, with a maximum deviation of $\sim$20 kcal/mol.
This large deviation suggests that DFT methods are not systematically accurate and there might be significant variation in the accuracy across various systems.
Efforts have been made to improve the DFT estimates by determining correction factors from fitting to experimental data~\cite{2013GRI}; however, these reduce the prediction ability of the theory.
Coupled cluster theory based methods such as CCSD(T) have been largely regarded as the ``gold standard'' for accuracy in quantum chemistry.
Several CCSD(T) studies have shown that the method can achieve sub-chemical accuracy if large enough basis sets and multiple corrections are used~\cite{2001FEL,2006FEL,2007FEL,2008FEL, 2012HAU}. 

\vspace{2mm}
QMC methods lie on the sweet spot of accuracy and computational cost.
They are far more accurate than mean-field methods (no inherent approximation like the XC functional) and have a more favorable scaling than the WF-based quantum chemical methods.
CCSD(T)/aug-cc-pVQZ gave an MAD of 2.8 kcal/mol for the G2 set atomization energies.
When extrapolated to the complete basis set limit, the MAD value was reduced to 1.3 kcal/mol.~\cite{1998FEL}
Several FN-DMC benchmark tests (using single SD \textit{ansatz}) for the G2 set atomization energies  have obtained MAD values close to 3 kcal/mol.
For instance, Nemec \textit{et al.}~\cite{2010NEM} used all-electron FN DMC on Slater determinant (SD) WF to obtain the atomization energies of the G2 set to an MAD of 3.2 kcal/mol.
Similar accuracy in atomization energies was also obtained previously in DMC pseudopotential calculations~\cite{2002GRO}. However, these simple DMC approaches (i.e., Jastrow + SD) have not been able to achieve chemical accuracy or sub-chemical accuracy due to the residual FN errors.

\vspace{2mm}
In this paper, we present FN DMC benchmark results using the so-called AGPs (antisymmetrized geminal power with singlet correlation) \textit{ansatz}~\cite{2003CAS} along with a Jastrow factor (JF). Combined with VMC optimization, the more flexible AGPs \textit{ansatz} leads to improved nodal surfaces.
The main outcome of this work is that the combination of a more flexible \textit{ansatz} (AGPs) and nodal surface optimization leads to a much better quality many body WF, which in turn leads to better DMC energies.
This is very important because, AGPs (even though being multiconfigurational in nature~\cite{2020NAK}) in practice is as efficient as a single determinant \textit{ansatz} and thus can be extended to much larger systems, even within the computationally demanding QMC methods.
QMC also provides an added advantage of the near ideal parallel scaling of QMC algorithms~\cite{2020NAK}.


\section{Computational details}
The {\sc{TurboRVB}}~\cite{2020NAK} QMC package was used for all calculations.
It employs resonating valence bond (RVB~\cite{1973AND}) WF and allows one to choose a more flexible \textit{ansatz} than the SD \textit{ansatz}, and includes correlation effects beyond the standard SD.

\subsection{Wavefunctions}
The choice of the WF \textit{ansatz} plays an important role in determining the accuracy and the computational cost of QMC calculations. A many-body WF \textit{ansatz} can be written as the following product:
\begin{equation}
    \Psi = \Phi_{{\rm{AS}}} * \exp J,
\end{equation}
where $\exp J$ is the JF, and $\Phi_{{\rm{AS}}}$ is the antisymmetric part that satisfies the antisymmetry condition for fermions. Generally, a single SD is used for the antisymmetric part in QMC calculations. SD is simply an antisymmetrized product of single particle electron orbitals and does not include any electron correlation by itself.

\subsubsection{Jastrow factor}
The Jastrow factor is a multiplier term which improves the quality of a many-body WF by providing a significant portion ($\approx 70\%$) of correlation energy and is necessary for fulfilling Kato's cusp conditions ~\cite{1957KAT}. The JF used here comprises three terms: one-body, two-body, and three/four body term ($J = J_1 + J_2 + J_{3/4}$). The one-body term is necessary to satisfy the electron-ion cusp condition. A separate one-body term is used for each element present in the molecule. It consists of the so-called homogeneous part,
\begin{equation}
    J_1^h(\textbf{r}_1, ..., \textbf{r}_N) = \sum_{i=1}^{N} \sum_{a=1}^{N_{at}} \left(-(2Z_a)^{3/4} u_a\left((2Z_a)^{1/4}|\textbf{r}_i - \textbf{R}_a|\right)\right), \label{hom}
\end{equation}
and the so-called inhomogeneous part,
\begin{equation}
    J_1^{inh}(\textbf{r}_1, ..., \textbf{r}_N) = \sum_{i=1}^{N} \sum_{a=1}^{N_{at}} \left( \sum_{l} M_{a, l} \chi_{a, l}(\textbf{r}_i) \right), \label{inhom}
\end{equation}
where, $\textbf{r}_i$ denotes the electron coordinates, $\textbf{R}_a$ represents the atomic positions, $Z_a$ denotes the corresponding atomic numbers, ${N_{at}}$ is the number of nuclei, $\chi_{a, l}$ represents a Gaussian-type atomic orbital $l$ centered over atom $a$, and $M_{a, l}$ denotes the corresponding variational parameters. The function $u_a$ is defined as:
\begin{equation}
    u_{a}(r) = \frac{1}{2b_{{\rm{e}}a}} \left( 1 - e^{-rb_{{\rm{e}}a}}\right), \label{u}
\end{equation}
where $b_{{\rm{e}}a}$ is a variational parameter that depends on each nucleus $a$.
The one-body terms were carefully optimized at the DFT level before final optimization at the VMC level. The two-body term is necessary to satisfy the electron-electron cusp condition and is optimized at the 
VMC level.
\begin{equation}
    J_2(\textbf{r}_1, ..., \textbf{r}_N) = \sum_{i<j} v (|\textbf{r}_i - \textbf{r}_j|). \label{j2}
\end{equation}
The function $v$ has the following form:
\begin{equation}
    v(r_{i, j}) = \frac{r_{i, j}}{2} (1 + b_{{\rm{ee}}}r_{i, j})^{-1}, \label{j2_d}
\end{equation}
where $r_{i, j} = |\textbf{r}_i - \textbf{r}_j|$ and $b_{{\rm{ee}}}$ is a single variational parameter. The three/four-body Jastrow term is defined as:
\begin{equation}
    J_{3/4}(\textbf{r}_1, ..., \textbf{r}_N) = \sum_{i<j} \left( \sum_{a, l} \sum_{b, m} M_{\{a, l\}, \{b, m\}} \chi_{a, l}(\textbf{r}_i), \chi_{b, m}(\textbf{r}_j) \right), \label{j3/4}
\end{equation}
where $M_{\{a, l\}, \{b, m\}}$ represents the variational parameters, $l$ and $m$ indicate orbitals centered on atomic sites $a$ and $b$, respectively. In the current work, the four-body Jastrow term (i.e., $a \neq b$) was not used.

\subsubsection{Antisymmetrized geminal power (AGP)}
One way to improve the description of electron correlations and move beyond the standard SD is to explicitly include pairwise correlation among electrons. This fairly general and flexible \textit{ansatz} is called as the Pfaffian WF ~\cite{2006BAJ}. 
The Pfaffian WF is constructed from pairing functions (known as geminals).
When only the singlet electron pairing terms are considered, we get the AGPs \textit{ansatz}~\cite{2003CAS}.
A generic pairing function for the AGPs, $f(\textbf{r}_i, \textbf{r}_j)$ can be written as:
\begin{equation}
    f(\textbf{r}_i, \textbf{r}_j) = \frac{1}{\sqrt{2}} (\ket{\uparrow \downarrow} - \ket{\downarrow \uparrow}) f_s(\textbf{r}_i, \textbf{r}_j). \label{agps_term}
\end{equation}
For a simpler unpolarized case, where the number of electrons $N$ is even and $N_\uparrow = N_\downarrow$, all possible combinations of singlet pairs can be written in the form of a matrix:
\begin{equation}
    F = \begin{psmallmatrix}
        f(\textbf{r}_{1}^{\uparrow}, \textbf{r}_{1}^{\downarrow}) & f(\textbf{r}_{1}^{\uparrow}, \textbf{r}_{2}^{\downarrow}) & \dots  & f(\textbf{r}_{1}^{\uparrow}, \textbf{r}_{N}^{\downarrow}) \\
        f(\textbf{r}_{2}^{\uparrow}, \textbf{r}_{1}^{\downarrow}) & f(\textbf{r}_{2}^{\uparrow}, \textbf{r}_{2}^{\downarrow}) & \dots  & f(\textbf{r}_{2}^{\uparrow}, \textbf{r}_{N}^{\downarrow}) \\
        \vdots                                                           & \vdots                                                           & \ddots & \vdots \\
        f(\textbf{r}_{N}^{\uparrow}, \textbf{r}_{1}^{\downarrow}) & f(\textbf{r}_{N}^{\uparrow}, \textbf{r}_{2}^{\downarrow}) & \dots & f(\textbf{r}_{N}^{\uparrow}, \textbf{r}_{N}^{\downarrow}) 
    \end{psmallmatrix}. \label{agps_matrix}
\end{equation}
The determinant of matrix \textit{F} gives the AGPs WF:
\begin{equation}
    \Phi_{{\rm{AGPs}}} = \det F \label{agps_eq}
\end{equation}
In case of an open-shell chemical system ($N_\uparrow \neq N_\downarrow$) the matrix $F$
consisting of singlet terms only, would become rectangular matrix.
To convert matrix $F$ into a square matrix, additional unpaired molecular orbitals ($\Theta_i(\textbf{r})$) are added to the matrix $F$.
Suppose, $N_\uparrow > N_\downarrow$, $N_\uparrow - N_\downarrow$ unpaired spin-up molecular
orbitals are added to the matrix $F$:

\begin{equation}
 F=    \begin{psmallmatrix}
 f(\textbf{r}_1^\uparrow, \textbf{r}_1^\downarrow) & f(\textbf{r}_1^\uparrow, \textbf{r}_2^\downarrow) & \cdots  & f(\textbf{r}_1^\uparrow, \textbf{r}_N^\downarrow) &                   \Theta_1(\textbf{r}_1) & \cdots  & \Theta_{N_\uparrow-N_\downarrow}(\textbf{r}_1) \\
 f(\textbf{r}_2^\uparrow, \textbf{r}_1^\downarrow) & f(\textbf{r}_2^\uparrow, \textbf{r}_2^\downarrow) & \cdots   & f(\textbf{r}_2^\uparrow, \textbf{r}_N^\downarrow) &                  \Theta_1(\textbf{r}_2) & \cdots  & \Theta_{N_\uparrow-N_\downarrow}(\textbf{r}_2) \\
  \vdots & \vdots  & \ddots  & \vdots  & \vdots  & \ddots  & \vdots \\
 f(\textbf{r}_N^\uparrow, \textbf{r}_1^\downarrow) & f(\textbf{r}_1^\uparrow, \textbf{r}_1^\downarrow) & \cdots  & f(\textbf{r}_N^\uparrow, \textbf{r}_N^\downarrow)  &                  \Theta_1(\textbf{r}_N) & \cdots   & \Theta_{N_\uparrow-N_\downarrow}(\textbf{r}_N)
 \end{psmallmatrix} \label{agps_matrix_polarized}
\end{equation}
$\det F$ now gives the antisymmetric WF.

\vspace{5mm}
The singlet electron pairing terms are represented by:
\begin{equation}
    f_s(\textbf{r}_i, \textbf{r}_j) = \sum_{a, l} \sum_{b, m} \lambda_{\{a,l\}, \{b,m\}} \phi_{a,l}(\textbf{r}_{i}) \phi_{b,m}(\textbf{r}_{j}), \label{geminal}
\end{equation}
where, $\phi_a$ and $\phi_b$ represent atomic orbitals centered on atoms $a$ and $b$, respectively; and the indices $l$ and $m$ indicate different orbitals centered on atoms $a$ and $b$, respectively. The elements of the matrix $\lambda$ are the coefficients or the variational parameters of the WF. 
An important advantage of the AGPs \textit{ansatz} is that it is equivalent to a linear combination of SDs (or multi-determinants) while maintaining the computational cost of a single determinant \textit{ansatz}~\cite{2020NAK}. 
Hence, this flexible \textit{ansatz} (greater variational freedom) could be an effective way to improve the quality of the many-body WF.


\subsection{Computational workflow}
The equilibrium geometries of the G2 set molecules were taken from previous benchmark studies~\cite{1997CUR,2008FEL,2005OEI} (see Table. S2 in supplementary information). 
The \textit{ansatz} was expanded using the triple-zeta atomic basis sets obtained from the Basis Set Exchange library \cite{2019PRI}. 
Larger exponents greater than $8Z^2$ (where \textit{Z} is the atomic number) were removed from the basis set to avoid numerical instabilities. 
The large exponent orbitals cut from the basis set are implicitly included by utilizing the one-body Jastrow term~\cite{2018MAZ,2019NAK}. 
The basis sets used for the determinant and Jastrow expansion are listed in Table.~\ref{basis_set_table}. 
The same basis sets were used for the VMC and LRDMC calculations. 


%
%

\begin{table}[htbp]
\centering
\caption{Basis set orbitals used for the determinant and Jastrow expansion}
\label{basis_set_table}
\begin{tabular}{@{}lcc@{}}
\toprule
Element & Det. basis       & Jas. basis   \\ \midrule
H       & 4s, 2p, 1d       & 4s, 2p       \\
Li      & 8s, 5p, 2d, 1f   & 8s, 5p, 2d   \\
C       & 8s, 5p, 2d, 1f   & 8s, 5p, 2d   \\
N       & 8s, 5p, 2d, 1f   & 8s, 5p, 2d   \\
O       & 7s, 5p, 2d, 1f   & 7s, 5p, 2d   \\
F       & 7s, 5p, 2d, 1f   & 7s, 5p, 2d   \\
Na      & 11s, 10p, 2d, 1f & 11s, 10p, 2d \\
Si      & 11s, 9p, 2d, 1f  & 11s, 9p, 2d  \\
P       & 11s, 9p, 2d, 1f  & 11s, 9p, 2d  \\
S       & 11s, 9p, 2d, 1f  & 11s, 9p, 2d  \\
Cl      & 11s, 9p, 2d, 1f  & 11s, 9p, 2d  \\ \bottomrule
\end{tabular}
\end{table}


The trial WF for the Jastrow single determinant \textit{ansatz} was obtained from DFT calculations using the {\sc{TurboRVB}} DFT module. To improve the efficiency of the DFT calculations, we utilized the double-grid DFT algorithm, which used a finer DFT mesh when in the vicinity of the nuclei \cite{2019NAK}. 
For the JDFT (JF + SD with DFT orbitals) \textit{ansatz}, the JF was optimized at the VMC level, which was followed by the LRDMC projection.
In LRDMC, instead of the conventional time discretization\cite{1993UMR} of the continuous Hamiltonian, the regularization of the original Hamiltonian is done over the lattice with a step size $a$, such that ${\hat{H}}^a \to \hat{H}$ for $a \to 0${~\cite{1994TEN, 1998BUO, 2000SOR}}.
Further details on the VMC and LRDMC algorithms can be found in Refs. {\citenum{2005CAS, 2006CAS, 2010CAS, 2017BEC, 2020NAK2}}.
The target error-bar for the DMC and VMC energies was taken as $\approx$0.3 mHa. 
For the VMC optimization, we used linear~\cite{2005SOR,2007TOU,2007UMR} and stochastic reconfiguration methods~\cite{1998SOR,2007SOR}.
The JDFT WF \textit{ansatz} was then converted into JsAGPs. 
No information loss occurs during this conversion because we are rewriting the SD \textit{ansatz} into a more flexible AGPs \textit{ansatz} and maximum overlap between the two WFs is ensured. 
The JsAGPs was then optimized at the VMC level, including both JF and nodal surface optimization. 
This was followed by the LRDMC projection and extrapolation to zero lattice space. 
The complexity of VMC optimization can be roughly estimated by the number of variational parameters to be optimized.
For instance, the number of variational parameters used for a simple system like BeH were 306 and 735 for JDFT and JsAGPs \textit{ansatz} respectively.
For a complex system like Si$_2$H$_6$ the number of variational parameters optimized were 872 and 6422 for JDFT and JsAGPs \textit{ansatz} respectively.
The variational energy, $E[\mathbf{\alpha}]$ as well as the maximum value of signal to noise ratio for forces (termed as \verb|devmax|) were monitored.
 
 \begin{equation}
     \verb| devmax | \equiv max \left ( \left | \frac{f_k}{\sigma_{f_k}} \right | \right ), \label{devmax}
 \end{equation}
where $f_k$ and $\sigma_{f_k}$ denote force and the corresponding error bar respectively.
It has been observed that when \verb|devmax| stabilizes to values $< \sim$4, an energy minimum with repect to the variational parameters is being approached~\cite{2020NAK}.
The same criteria was used (in addition to checking a general convergence of energy and one-/two-body Jastrow       parameters) in general to decide when to stop VMC optimization.
For instance, for Si$_2$H$_6$ VMC optimization was carried out for 4000 optimization steps using linear method.
For most of the molecules considered, the linear method was used for VMC optimization.
For SiH$_3$, C$_2$H$_2$, CH$_2$ (triplet) and CH$_3$ the stochastic method was used, requiring
$\sim$15000-20000 optimization steps.

\vspace{5mm}

Majority of the calculations were performed on the supercomputer Fugaku using 2304 CPU cores distributed across 48 nodes. 
To improve the efficiency of the calculations, we used {\sc{TurboGenius}}, a python-based wrapper for {\sc{TurboRVB}}, which is useful in performing high-throughput calculations \cite{2020NAK}.


\section{Results}
\subsection{JsAGPs for N$_2$ molecule}
To validate the methodology and verify whether the basis sets used were good enough to approach chemical accuracy, the energies of N$_2$ molecule and N atom were compared with previous benchmark tests and experimental values (see Table.~{\ref{n_energies_table}}).
\begin{table*}[htbp]
\caption{Nitrogen energies}
\label{n_energies_table}
\begin{tabular}{@{}lccccc@{}}
\toprule
 & Method                                                          & Basis set & Atom (Ha)    & Molecule (Ha) & Atomization energy (eV) \\ \midrule
 & JDFT-VMC                                                        & cc-pVTZ  & $-54.5543(2)$  & $-109.4522(3)$  & $9.35(1)$      \\
 &
  {\color[HTML]{9B9B9B} JDFT-DMC} &
  {\color[HTML]{9B9B9B} cc-pVTZ} &
  {\color[HTML]{9B9B9B} $-54.5765(3)$} &
  {\color[HTML]{9B9B9B} $-109.5068(3)$} &
  {\color[HTML]{9B9B9B} $9.61(2)$} \\
 & JsAGPs-VMC                                                       & cc-pVTZ  & $-54.5614(1)$  & $-109.4702(5)$  & $9.45(1)$      \\
\multirow{-4}{*}{\begin{tabular}[c]{@{}l@{}}Current work\\ ({\sc{TurboRVB}})\end{tabular}} &
  {\color[HTML]{9B9B9B} JsAGPs-DMC} &
  {\color[HTML]{9B9B9B} cc-pVTZ} &
  {\color[HTML]{9B9B9B} $-54.5785(4)$} &
  {\color[HTML]{9B9B9B} $-109.5165(3)$} &
  {\color[HTML]{9B9B9B} $9.92(1)$} \\ \midrule
    & \begin{tabular}[c]{@{}c@{}}JHF-DMC$^a$\end{tabular} & QZ4P  & $-54.5765(2)$  & $-109.5065(4)$  & $9.61(2)$      \\
 &
  {\color[HTML]{9B9B9B} CCSD(T)$^b$} &
  {\color[HTML]{9B9B9B} --} &
  {\color[HTML]{9B9B9B} --} &
  {\color[HTML]{9B9B9B} --} &
  {\color[HTML]{9B9B9B} $9.85(1)$} \\
 & Fermi net$^c$                                                   & --        & $-54.58882(6)$ & $-109.5388(1)$  & $9.828(5)$     \\
 &
    {\color[HTML]{9B9B9B} \begin{tabular}[c]{@{}c@{}}JSD-DMC\\ pseudopotential$^d$\end{tabular}} &
  {\color[HTML]{9B9B9B} cc-pV5Z} &
  {\color[HTML]{9B9B9B} --} &
  {\color[HTML]{9B9B9B} --} &
  {\color[HTML]{9B9B9B} $9.573(4)$} \\
 & Estimated exact$^e$                                             & --        & $-54.5892$     & $-109.5427$     & $9.91$         \\
\multirow{-6}{*}{Previous reports} &
  {\color[HTML]{9B9B9B} Experimental$^f$} &
  {\color[HTML]{9B9B9B} --} &
  {\color[HTML]{9B9B9B} --} &
  {\color[HTML]{9B9B9B} --} &
  {\color[HTML]{9B9B9B} $9.91$} \\ \bottomrule
\end{tabular}

\vspace{2mm}
\textsuperscript{\emph{a}} Reference{~\citenum{2010NEM}};
\textsuperscript{\emph{b}} Reference{~\citenum{2008FEL}};
\textsuperscript{\emph{c}} Reference{~\citenum{2020PFA}};
\textsuperscript{\emph{d}} Reference{~\citenum{2012PET}};
\textsuperscript{\emph{e}} Reference{~\citenum{2005BYT}};
\textsuperscript{\emph{f}} Reference{~\citenum{1999FEL}}.
\end{table*}
 
The JsAGPs-DMC atomization energy shows excellent agreement with the experimental value~\cite{1999FEL}. 
It is better than the value computed using CCSD(T)~\cite{2008FEL} and the value computed recently using a neural network based \textit{ansatz} (called as Fermi net)~\cite{2020PFA}.
The correlation energies recovered for the N$_2$ molecule at the JsAGPs-DMC level and the JDFT-DMC level were $\approx$95\% and $\approx$93\% respectively.
At the VMC level, for JsAGPs, $\approx$86\% percent correlation energy was recovered.
The computed JDFT-DMC energies are in excellent agreement with the ones computed by Nemec \textit{et al.}~\cite{2010NEM}.
Clearly, in the case of nitrogen, a triple-zeta basis set for orbital and Jastrow expansion was good enough.
The best total (closest to the estimated exact) energies are the ones computed by Pfau \textit{et al.} using Fermi net~\cite{2020PFA}. The JsAGPs atomization energy however, is more accurate than that obtained by Pfau \textit{et al.}.
This shows that the JsAGPs \textit{ansatz} allows remarkable cancellation of errors when the difference between the molecular and atomic energies is computed.
Interestingly, the N$_2$ atomization energy computed using the Jastrow Slater determinant (JSD) \textit{ansatz} by Petruzielo \textit{et al.}~\cite{2012PET} could not approach CCSD(T) level accuracy. 
Hence, for the JSD \textit{ansatz}, Jastrow and nodal surface optimizations are not sufficient for improving the quality of the WF \textit{ansatz}. 
Thus, the standard JSD might be inadequate for this purpose.


\subsection{Application to the G2 set}

\begin{figure*}[]
\centering
 \includegraphics[width=0.85\linewidth]{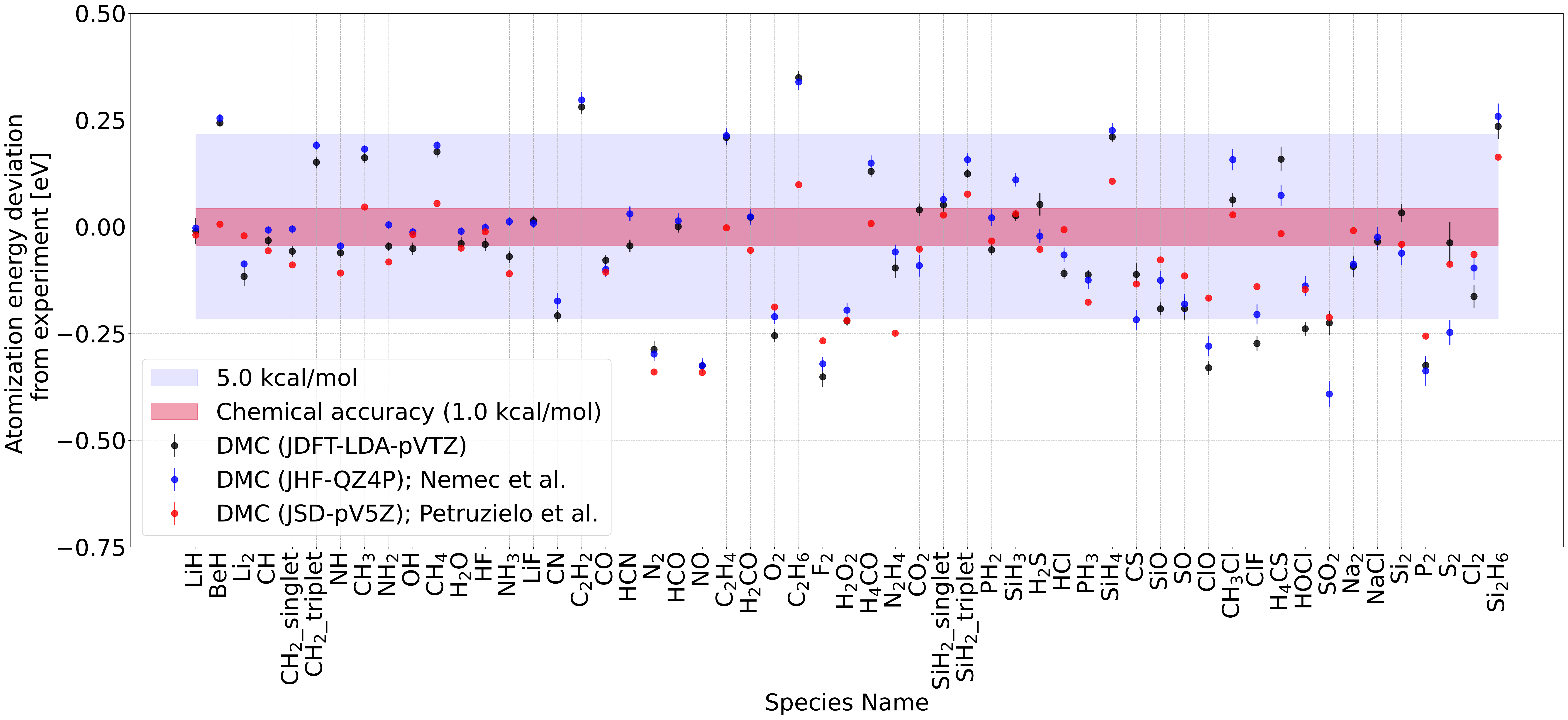}
    \caption{Deviation of the DMC atomization energies from the experimentally obtained values for the JDFT \textit{ansatz} with the triple zeta basis set. Zero point energies and relativistic + spin orbit were corrected before computing the deviations between the DMC and experimental values~\cite{2010NEM}. Values obtained by Nemec \textit{et al.}~\cite{2010NEM} (JHF-QZ4P) and Petruzielo \textit{et al.}~\cite{2012PET} (JSD-5z) are also plotted for comparison. The MAD for the JDFT \textit{ansatz} is $\approx$3.2 kcal/mol and the MAD values obtained by Nemec \textit{et al.} and Petruzielo \textit{et al.} are 3.13 kcal/mol and 2.1 kcal/mol, respectively.}
 \label{Figure1}
\end{figure*}

Our JDFT-DMC atomization energies (Fig.~\ref{Figure1}) were first compared with the ones obtained by Nemec \textit{et al.}~\cite{2010NEM}(Jastrow Hartee Fock (JHF)-DMC, QZ4P STO basis set) and Petruzielo \textit{et al.}~\cite{2012PET}(JSD-DMC, 5z basis set).
The results obtained for the JDFT-DMC atomization energies are in good agreement with the ones obtained by Nemec \textit{et al.}
Most atomization energies obtained using the JDFT \textit{ansatz} were within a deviation of $\pm$0.25 eV ($\pm$3.0 kcal/mol) from the experimental values although very few were in the chemical accuracy range.
The JDFT atomization energies had an MAD of $\approx$3.2 kcal/mol, which is quite close to the value of $\approx$3.13 kcal/mol reported by Nemec \textit{et al.}
An MAD of 2.9 kcal/mol was reported in another FN DMC (atomic cores treated with pseudopotentials) G2 set benchmark by Grossman~\cite{2002GRO}.
This overall agreement with previous benchmark tests points out that FN DMC provides ``near chemical accuracy'' and the primary sources of error are the fixed (unoptimized) nodes.
There could be other sources of error, such as the basis set used for orbital expansion. However, it can be ruled out based on the fact that Nemec \textit{et al.} used a larger basis set (Qz) than that used in the current study.
Nonetheless, the errors in the atomization energies are quite similar.
Comparison of the JDFT atomization energies with the JSD atomization energies (obtained by Petruzielo \textit{et al.}) shows that optimizing the nodal surfaces improves the DMC atomization energy estimates over the ones obtained using DFT or mean-field nodal surfaces.

\begin{figure*}[]
\centering
\includegraphics[width=0.85\linewidth]{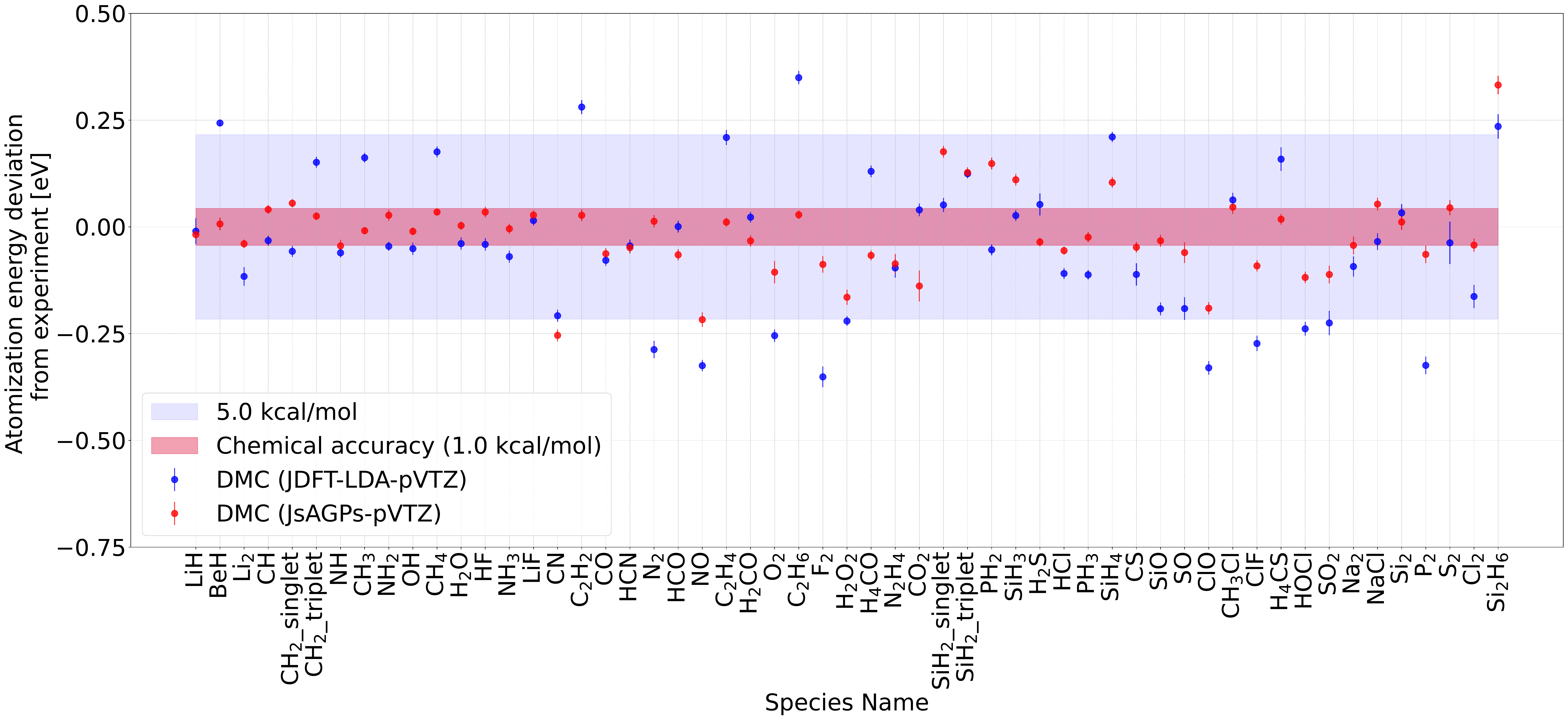}
    \caption{Deviation of the DMC atomization energies from the experimentally obtained values for the JDFT and JsAGPs \textit{ansatz}. The error-bars are shown within the markers. The MAD from the experiment for the JDFT and JsAGPs \textit{ansatz} are $\approx$3.2 kcal/mol and $\approx$1.6 kcal/mol, respectively. Zero point energies and relativistic + spin orbit were corrected before computing the deviations between the DMC and experimental values~\cite{2010NEM}.}
\label{Figure2}
\end{figure*}


\begin{figure*}[]
\centering
 \includegraphics[width=\linewidth]{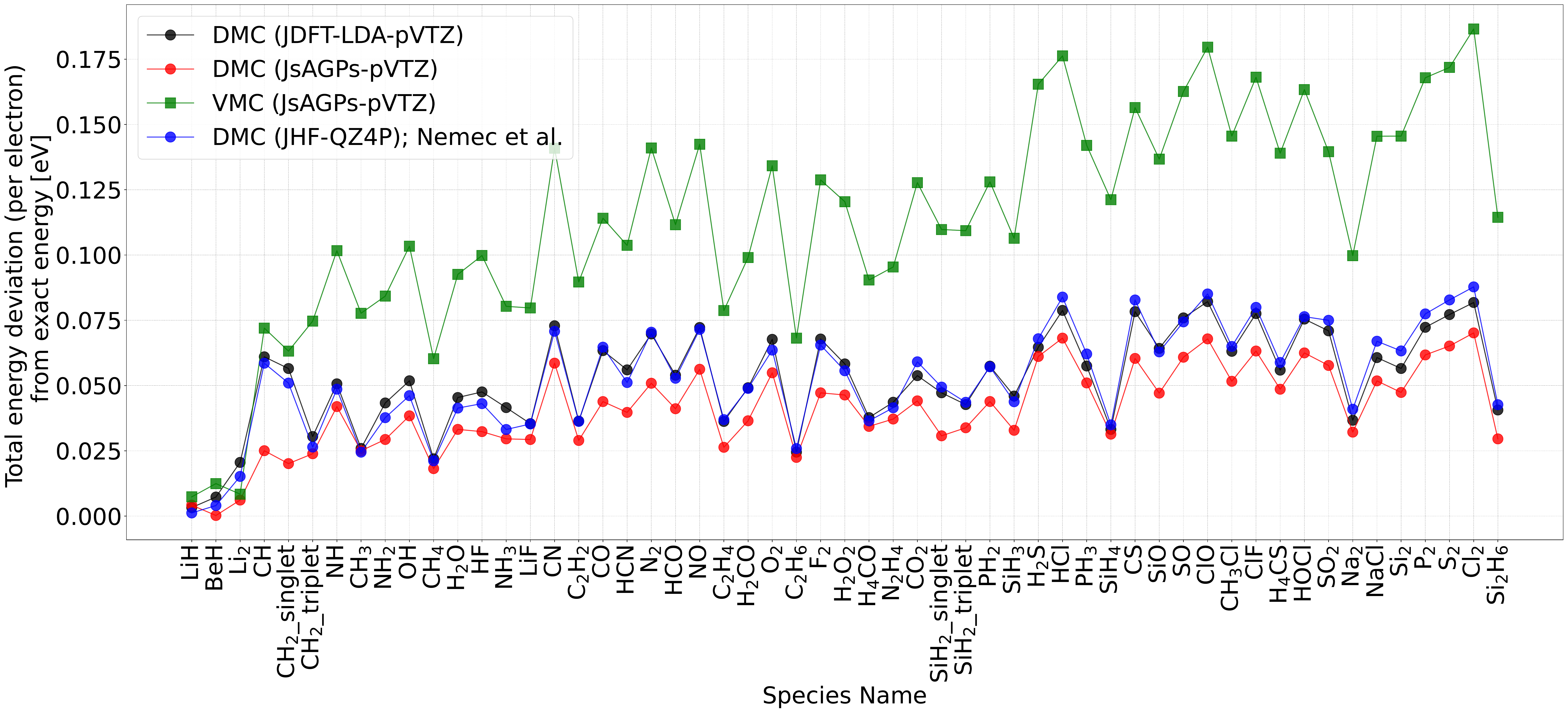}
    \caption{Deviation of the VMC and DMC total energies from the estimated exact energies. Values obtained by Nemec \textit{et al.}~\cite{2010NEM} are plotted for comparison.}
 \label{Figure3}
\end{figure*}

\begin{figure*}[]
\centering
 \includegraphics[width=\linewidth]{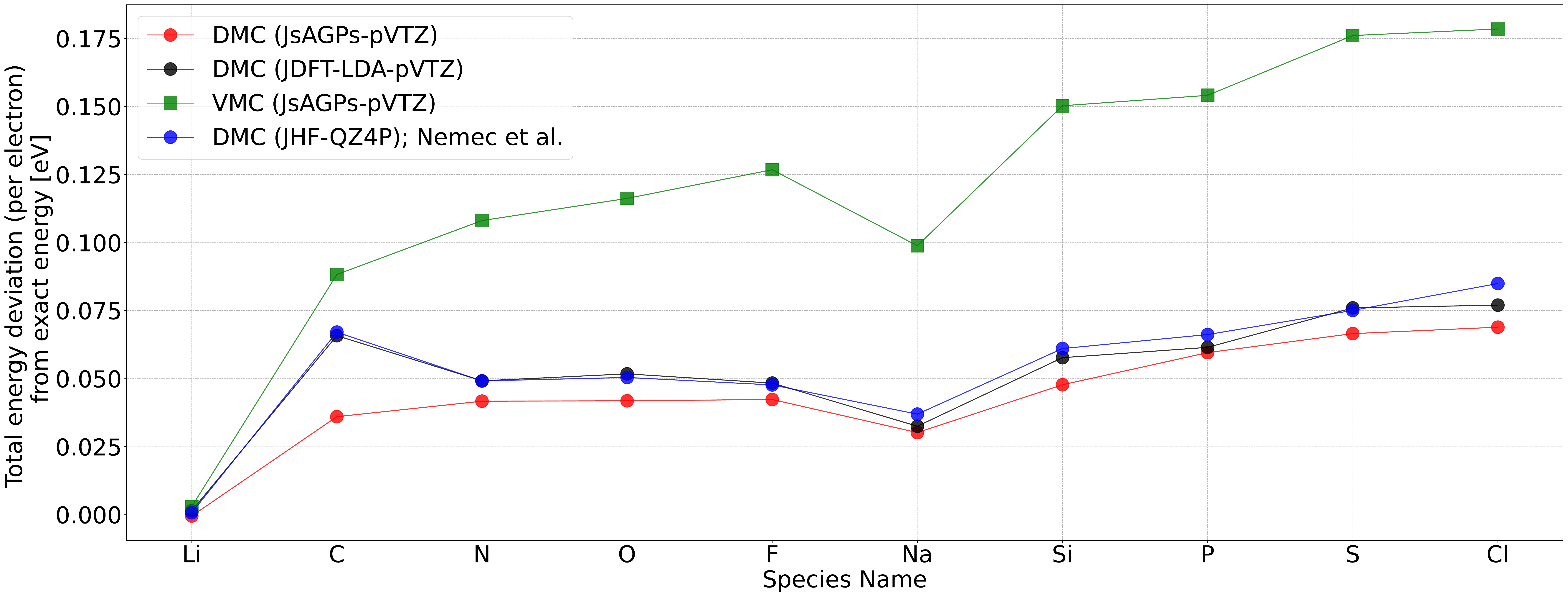}
    \caption{Deviation of the VMC and DMC total energies of atoms from the estimated exact energy. Values obtained by Nemec \textit{et al.}~\cite{2010NEM} are plotted for comparison.}
 \label{Figure4}
\end{figure*}

AGPs, which is a more flexible \textit{ansatz}, allows for better nodal surfaces.
Fig.~{\ref{Figure2}} shows a comparison of the DMC atomization energy between the JDFT and the more flexible JsAGPs
\textit{ansatz}.
Total energies are shown in Fig.~\ref{Figure3} and Fig.~\ref{Figure4}.
Clearly, the best variational energies were obtained when LRDMC was applied to the JsAGPs \textit{ansatz} both for the atoms and the molecules. 
This means that the nodal surfaces of JsAGPs WF are better than those of JDFT, and hence, considerably more correlation energy is recovered.
For example, in case of atoms, an average of 95.6\% correlation energy was recovered at the DMC level for the JsAGPs WF and 93.7\% correlation energy was recovered using the JDFT.
In case of atomization energies, JsAGPs is clearly better than JDFT. 
The MAD using the JsAGPs and JDFT \textit{ansatz} were 1.6 kcal/mol and 3.2 kcal/mol, respectively, demonstrating a clear superiority of the JsAGPs. 
For almost all the molecules, the error was within 5.0 kcal/mol, and chemical accuracy was achieved for 26 molecules in the G2 set. 
These results are not only better than the ones obtained using JDFT (with nodal surface from DFT) but also better than the case wherein the SD nodal surfaces were optimized by Petruzielo \textit{et al.}~\cite{2012PET}. 
The better atomization energies indicate that JsAGPs not only provides better variational energies but also improves error cancellation.

\section{Discussion}
It is interesting to note that for a few molecules (eg. Si$_2$H$_6$, CO$_2$), the JsAGPs atomization energies are worse than their JDFT counterparts. It turns out that although AGPs leads to better nodal surfaces (and lower DMC energies) than JDFT for all molecules and atoms, the quality of nodal surface depends upon the chemical structure of molecules and atoms, and hence, the error cancellation is not always predictably better.

\vspace{2mm}
In this study, we tried to reduce the FN error by optimizing the nodal surfaces at the variational level with the single-determinant \textit{ansatz} (JsAGPs). It is important to compare our QMC results with those obtained with the multi-determinant \textit{ansatz} that is promising in terms of accuracy.
Petruzielo \textit{et al.}~\cite{2012PET} showed that DMC with the multi-determinant ansatz constructed from multi-configurational self-consistent field theory (MCSCF) calculations achieves the MAD of 1.2 kcal/mol.
Morales \textit{et al.}~\cite{2012MOR} were able to achieve sub-chemical accuracy (MAD of 0.8 kcal/mol) in atomization energies of the G2 set.
In another study, chemical accuracy (and almost exact energies) could be achieved for the ionization potentials for several atoms from Li to Mg using full-configuration interaction~\cite{2010BOO}. 
Yao \textit{et al.} used the recently developed semistochastic heat-bath configuration interaction method to obtain excellent atomization energies for the G2 set with an MAD of 0.46 kcal/mol~\cite{2020YAO}. 
Scemama \textit{et al.}~\cite{2020SCE} have recently developed a way to combine short-range XC functionals (from DFT) with selected configuration interaction, which led to trial WFs with lower FN energies with compact multi-determinant trial wave functions. 
The least MAD obtained reached 2.06 kcal/mol.
Thus, the MADs obtained with the multi-determinant approach are compatible with or better than the value obtained using the JsAGPs \textit{ansatz} in this study.

\vspace{2mm}
\noindent
The advantage of the multi-determinant approach is that one can systematically improve the accuracy by increasing the number of SDs, and, in theory, it can describe any ground state exactly with a sufficient number of SDs~\cite{2020GEN}.
However, the number of SDs that should be considered scales exponentially as the system size increases, which makes the approach computationally demanding for larger systems.
While, in recent years, there have been successful efforts to reduce the number of determinants required for accurate calculations by truncating less important determinants~{\cite{2015GIN, 2016MIC}}, establishing a technique to apply the multi-determinant \textit{ansatz} to large systems with high accuracy is still an active field of research~{\cite{2020SCE, 2020BEN}}.
Instead, the single-determinant approach we employed in this study enables the \textit{ansatz} to have a large variational freedom, while keeping the computational cost lower, which allows one to tackle large molecules~\cite{2020GEN}. It is an advantage of our single-determinant approach from the practical viewpoint.
However, it is worth mentioning that a flexible single-determinant \textit{ansatz} like the AGPs might suffer from its own limitations. For instance, the \textit{ansatz} cannot be improved systematically unlike in the multi-determinant approach.
Thus, one should devise an appropriate \textit{ansatz} from the physics viewpoint. For instance, Claudio \textit{et al.}~{\cite{2020GEN, 2020GEN2}} demonstrated that the atomization energy of the carbon dimer, for which spin fluctuations should be considered, is significantly improved by using the Pfaffian \textit{ansatz}{~\cite{2006BAJ}}.
Another limitation is that the optimization of a large number of nonlinear variational parameters in the single determinant \textit{ansatz} should be handled efficiently. More specifically, due to the large number of variational parameters contained in the single determinant \textit{ansatz}, the optimization may fall into physically incorrect local minima, or the optimization itself may diverge. To prevent them, in practice, one needs to prepare a good initial estimate and/or reduce the number of variational parameters by applying constraints in a physically meaningful way, but there are no general guidelines for these.
Thus, applying QMC methods to large-scale systems beyond the fixed-node obtained by DFT or HF still requires developing new techniques.



\section{Conclusions}
To conclude, we have demonstrated the effectiveness of the AGPs ~\textit{ansatz} (built from electron pairing functions or geminals) in VMC and LRDMC calculations using the {\sc{TurboRVB}} QMC package.
Using AGPs \textit{ansatz} with the cc-pVTZ basis set, the LRDMC calculations for atoms recovered 95.6\% correlation energy.
The atomization energies computed using AGPs had an MAD of $\sim$1.6 kcal/mol from the experimental values.
Chemical accuracy was achieved for several molecules and the error was within $\pm$ 5 kcal/mol for almost all the molecules.
These results are quite encouraging as they show that allowing more variational freedom by utilizing more flexible \textit{ansatz} is a viable path towards more accurate QMC calculations.
We believe that our work represents an important step towards showing that a combination of more flexible single determinant \textit{ansatz} (JsAGPs) and nodal surface optimization can match the accuracy of quantum chemistry calculations, with the added advantages of excellent scaling and lower computational cost. 

\section{Conflict of interest}
The authors declare no conflict of interest.

\section{Code availability} \label{sec:code}
{\sc{TurboRVB}} is available from the website [https://turborvb.sissa.it] upon request.

\begin{suppinfo}

The following files are available free of charge.
\begin{itemize}
  \item supplement.pdf: Computed total energies and references for molecular geometries 
\end{itemize}

\end{suppinfo}


\begin{acknowledgement}
The computations in this work were mainly performed using the supercomputer Fugaku provided by RIKEN through the HPCI System Research Projects (Project IDs: hp210038 and hp220060). 
The authors are also grateful for the computational resources at the Research Center for Advanced Computing Infrastructure at Japan Advanced Institute of Science and Technology (JAIST).
A.R. is grateful to MEXT Japan for the support through the MEXT scholarship.
R.M. is grateful for financial support from MEXT-KAKENHI (22H05146, 21K03400 and 19H04692) and from JSPS Bilateral Joint Projects (JPJSBP120197714).
K.H. is grateful for the financial support from MEXT-KAKENHI, Japan (JP19K05029, JP21K03400, JP21H01998, and JP22H02170).
K.N. acknowledges financial support from the JSPS Overseas Research Fellowships, from Grant-in-Aid for Early Career Scientists (Grant No. JP21K17752), from Grant-in-Aid for Scientific Research (Grant No. JP21K03400), and from MEXT Leading Initiative for Excellent Young Researchers (Grant No. JPMXS0320220025).
This work was partly supported by the European Centre of Excellence in Exascale Computing (TREX-Targeting Real Chemical Accuracy at the Exascale). This project received funding from the European Union's Horizon 2020 Research and Innovation program under Grant Agreement No. 952165.
Finally, we dedicate this paper to the memory of Prof. Sandro Sorella (SISSA), who passed away during this collaboration. We remember him as one of the most influential contributors to the QMC community in the past century, and in particular for deeply inspiring this work with the development of the ab initio QMC code, {\sc{TurboRVB}}.
\end{acknowledgement}

\clearpage
\twocolumn

\bibliography{./references.bib}

\end{document}


\begin{center}
    \LARGE \textbf{Supplementary Information}
\end{center}

\section{Computed total energies using various \textit{ansatz}}
Total energies of atoms and molecules computed using various \textit{ansatz} are listed in the tables below.
For, molecules the geometry references are also provided.

%
%

{\footnotesize \begin{table}[htbp]
\centering
\caption{Total energies [Ha] for atoms}
\label{atom_energies_table}
\begin{tabular}{lllllr}
\toprule
    Element & DMC-JsAGPs & DMC-JDFT & VMC-JsAGPs & DMC-JHF$^+$ &  Est. exact$^*$ \\
\midrule
          Li &       -7.47815(7) &       -7.4779(2) &       -7.47777(6) &       -7.47802(6) &      -7.4781 \\
           C &       -37.8371(2) &      -37.8305(3) &       -37.8255(2) &       -37.8302(3) &     -37.8450 \\
           N &       -54.5785(4) &      -54.5765(3) &       -54.5614(1) &       -54.5766(2) &     -54.5892 \\
           O &       -75.0550(3) &      -75.0521(2) &       -75.0331(2) &       -75.0525(2) &     -75.0673 \\
           F &       -99.7199(3) &      -99.7179(2) &       -99.6920(2) &       -99.7181(2) &     -99.7339 \\
          Na &      -162.2424(3) &     -162.2415(3) &      -162.2146(2) &      -162.2397(2) &    -162.2546 \\
          Si &      -289.3344(3) &     -289.3293(3) &      -289.2817(2) &      -289.3276(4) &    -289.3590 \\
           P &      -341.2262(3) &     -341.2251(3) &      -341.1740(2) &      -341.2225(6) &    -341.2590 \\
           S &      -398.0709(3) &     -398.0653(9) &      -398.0065(2) &      -398.0659(4) &    -398.1100 \\
          Cl &      -460.1050(2) &     -460.0999(4) &      -460.0365(2) &      -460.0949(4) &    -460.1480 \\
\bottomrule
\end{tabular}

\end{table}}

%
%

{\footnotesize \begin{longtable} [htbp] {llllllr}
\caption{Total energies [Ha] for molecules}
\label{molecular_energies_table}\\
\toprule
    species\_list & Geometry & DMC-JsAGPs & DMC-JDFT & VMC-JsAGPs & DMC-JHF$^+$ &  Est. exact$^*$ \\
\midrule
\endfirsthead
\caption[]{Total energies [Ha] for molecules} \\
\toprule
    species\_list & Geometry & DMC-JsAGPs & DMC-JDFT & VMC-JsAGPs & DMC-JHF$^+$ &  Est. exact$^*$ \\
\midrule
\endhead
\midrule
\multicolumn{7}{r}{{Continued on next page}} \\
\midrule
\endfoot

\bottomrule
\endlastfoot
            LiH &  a&      -8.0699(3) &        -8.070(1) &       -8.06945(7) &        -8.0704(2) &    -8.070529 \\
            BeH &  a&     -15.2467(5) &      -15.2454(2) &       -15.2445(2) &       -15.2460(2) &   -15.246761 \\
         Li$_2$ &  a&     -14.9937(3) &      -14.9905(7) &       -14.9932(2) &       -14.9917(2) &   -14.995084 \\
             CH &  a&     -38.4724(3) &      -38.4632(3) &       -38.4603(3) &       -38.4638(3) &   -38.478863 \\
 CH$_2$\_singlet & a&      -39.1280(3) &      -39.1173(4) &      -39.11534(7) &       -39.1189(3) &   -39.133920 \\
 CH$_2$\_triplet & a&      -39.1420(3) &      -39.1401(4) &       -39.1271(2) &       -39.1413(2) &   -39.149059 \\
             NH &  a&     -55.2104(3) &      -55.2078(2) &       -55.1929(1) &       -55.2085(3) &   -55.222744 \\
         CH$_3$ &  a&     -39.8276(3) &      -39.8273(3) &       -39.8101(1) &       -39.8277(2) &   -39.835829 \\
         NH$_2$ &  a&     -55.8698(3) &      -55.8652(2) &       -55.8517(1) &       -55.8671(3) &   -55.879554 \\
             OH &  a&     -75.7248(1) &      -75.7204(5) &       -75.7033(2) &       -75.7222(3) &   -75.737497 \\
         CH$_4$ &  a&     -40.5086(2) &      -40.5072(4) &      -40.49311(8) &       -40.5075(3) &   -40.515269 \\
         H$_2$O &  a&     -76.4270(2) &      -76.4225(5) &       -76.4052(2) &       -76.4240(3) &   -76.439167 \\
             HF &  a&     -100.4478(3) &     -100.4422(5) &      -100.4230(2) &      -100.4439(3) &  -100.459713 \\
         NH$_3$ &  a&     -56.5539(2) &      -56.5495(4) &      -56.53521(9) &       -56.5525(3) &   -56.564731 \\
            LiF &  a&    -107.4214(3) &     -107.4187(3) &      -107.3991(2) &      -107.4187(2) &  -107.434307 \\
             CN &  b&     -92.6950(3) &      -92.6881(3) &       -92.6556(1) &       -92.6891(4) &   -92.722961 \\
     C$_2$H$_2$ &  a&     -77.3175(3) &      -77.3137(3) &       -77.2862(2) &       -77.3137(4) &   -77.332381 \\
             CO &  b&    -113.3038(3) &     -113.2937(3) &      -113.2676(2) &      -113.2930(4) &  -113.326318 \\
            HCN &  a&     -93.4107(3) &      -93.4023(3) &       -93.3777(2) &       -93.4048(4) &   -93.431085 \\
          N$_2$ &  b&    -109.5165(3) &     -109.5068(3) &      -109.4702(5) &      -109.5065(4) &  -109.542697 \\
            HCO &  a&    -113.8342(3) &     -113.8272(4) &      -113.7954(2) &      -113.8278(4) &  -113.856915 \\
             NO &  b&    -129.8696(3) &     -129.8608(3) &     -129.82208(8) &      -129.8612(4) &  -129.900576 \\
     C$_2$H$_4$ &  a&     -78.5735(2) &      -78.5676(4) &       -78.5427(2) &       -78.5672(4) &   -78.588951 \\
        H$_2$CO &  a&    -114.4876(3) &     -114.4802(3) &      -114.4509(2) &      -114.4803(4) &  -114.509104 \\
          O$_2$ &  b&    -150.2949(3) &     -150.2874(4) &      -150.2483(2) &      -150.2898(4) &  -150.327203 \\
     C$_2$H$_6$ &  a&     -79.8120(1) &      -79.8107(2) &       -79.7818(2) &       -79.8098(5) &   -79.826875 \\
          F$_2$ &  a&    -199.4989(3) &     -199.4852(8) &      -199.4449(2) &      -199.4868(4) &  -199.530110 \\
     H$_2$O$_2$ &  a&   -151.53355(9) &     -151.5257(2) &      -151.4846(2) &      -151.5274(4) &  -151.564235 \\
        H$_4$CO &  a&    -115.7081(2) &     -115.7058(4) &      -115.6710(2) &      -115.7066(4) &  -115.730774 \\
     N$_2$H$_4$ &  a&    -111.8531(3) &     -111.8488(5) &      -111.8146(2) &      -111.8502(5) &  -111.877672 \\
         CO$_2$ &  b&    -188.5658(3) &     -188.5580(3) &      -188.4982(2) &      -188.5537(6) &  -188.601471 \\
SiH$_2$\_singlet & c&     -290.5836(4) &     -290.5739(5) &      -290.5372(2) &      -290.5727(4) &  -290.601705 \\
SiH$_2$\_triplet & a&     -290.5490(3) &     -290.5438(2) &     -290.50460(9) &      -290.5433(4) &  -290.568877 \\
         PH$_2$ &  c&    -342.4759(4) &     -342.4674(3) &      -342.4233(2) &      -342.4676(5) &  -342.503299 \\
        SiH$_3$ &  a&    -291.2015(4) &     -291.1933(3) &      -291.1556(2) &      -291.1947(4) &  -291.222022 \\
         H$_2$S &  a&    -399.3620(2) &     -399.3596(3) &      -399.2930(2) &      -399.3575(4) &  -399.402426 \\
            HCl &  a&    -460.7743(2) &     -460.7673(2) &      -460.7028(2) &      -460.7639(5) &  -460.819376 \\
         PH$_3$ &  a&    -343.1141(3) &     -343.1098(2) &      -343.0539(2) &      -343.1068(5) &  -343.147839 \\
        SiH$_4$ &  a&    -291.8530(3) &     -291.8518(3) &      -291.7936(2) &      -291.8506(5) &  -291.873733 \\
             CS &  a&    -436.1801(3) &     -436.1656(2) &      -436.1024(2) &      -436.1620(6) &  -436.228940 \\
            SiO &  a&    -364.6950(3) &     -364.6811(4) &      -364.6225(2) &      -364.6822(5) &  -364.733068 \\
             SO &  b&    -473.3246(8) &     -473.3113(3) &      -473.2348(2) &      -473.3126(6) &  -473.378253 \\
            ClO &  b&    -535.2580(4) &     -535.2449(4) &      -535.1553(2) &      -535.2422(6) &  -535.320350 \\
       CH$_3$Cl &  a&    -500.0746(5) &     -500.0636(4) &      -499.9848(2) &      -500.0618(6) &  -500.123907 \\
            ClF &  a&    -559.9217(3) &     -559.9080(5) &      -559.8215(2) &      -559.9057(6) &  -559.982138 \\
        H$_4$CS &  a&    -438.6654(3) &     -438.6584(4) &      -438.5790(2) &      -438.6556(6) &  -438.711801 \\
           HOCl &  a&    -535.9203(3) &     -535.9079(4) &      -535.8239(2) &      -535.9070(6) &  -535.979997 \\
         SO$_2$ &  a&    -548.5908(3) &     -548.5752(4) &      -548.4945(2) &      -548.5704(7) &  -548.658618 \\
         Na$_2$ &  c&    -324.5103(4) &     -324.5066(6) &      -324.4556(2) &      -324.5032(4) &  -324.536291 \\
           NaCl &  a&    -622.5069(4) &     -622.4977(5) &      -622.4105(2) &      -622.4913(6) &  -622.560207 \\
         Si$_2$ &  a&    -578.7899(3) &     -578.7805(3) &      -578.6889(2) &      -578.7735(6) &  -578.838636 \\
          P$_2$ &  a&    -682.6364(4) &     -682.6247(3) &      -682.5193(2) &      -682.6191(7) &  -682.704451 \\
          S$_2$ &  a&    -796.3076(4) &     -796.2935(3) &      -796.1822(2) &      -796.2869(8) &  -796.384237 \\
         Cl$_2$ &  a&    -920.3023(3) &     -920.2878(6) &      -920.1569(2) &      -920.2803(6) &  -920.389991 \\
    Si$_2$H$_6$ &  c&    -582.5298(5) &     -582.5160(8) &      -582.4238(2) &      -582.5134(8) &  -582.566752 \\
\end{longtable}}

\begin{footnotesize}
$^+$Nemec \textit{et al.}, J. Chem. Phys. 2010, 132, 034111

$^*$Bytautas \textit{et al.}, J. Chem. Phys. 2005, 122, 154110

$^a$Curtiss \textit{et al.}, J. Chem. Phys. 1997, 106, 1063–1079

$^b$Feller \textit{et al.}, J. Chem. Phys. 2008, 129, 204105

$^c$O'eill \textit{et al.}, Mol. Phys. 2005, 103, 763-766

\end{footnotesize}